\documentclass[prl,superscriptaddress,twocolumn]{revtex4-1}
\usepackage{times,amsmath,upgreek}
\usepackage{epsfig}
\usepackage{graphicx}
\usepackage{dcolumn}
\usepackage{bm}
\usepackage{tabularx}
\usepackage{hyperref}
\hypersetup{citecolor=black, filecolor= black, linkcolor= black, urlcolor= black}
\begin{document}

\title{Constrained evolutionary algorithm for structure prediction of molecular crystals: methodology and applications}
\author{Qiang Zhu}
\email{qiang.zhu@stonybrook.edu}
\affiliation{Department of Geosciences, Department of Physics and Astronomy, and New York Center for Computational Sciences, Stony Brook University, Stony Brook, New York 11794, USA}
\author{Artem R. Oganov}
\affiliation{Department of Geosciences, Department of Physics and Astronomy, and New York Center for Computational Sciences, Stony Brook University, Stony Brook, New York 11794, USA}
\affiliation{Geology Department, Moscow State University, 119992, Moscow, Russia}
\author{Colin W. Glass}
\affiliation{High Performance Computing Center Stuttgart (HLRS), Germany}
\author{Harold T. Stokes}
\affiliation{Department of Physics and Astronomy, Brigham Young University, Provo, Utah 84602, USA}

\begin{abstract}
Evolutionary crystal structure prediction proved to be a powerful approach for studying a wide range of materials. Here, we present a specifically designed algorithm for the prediction of the structure of complex crystals consisting of well-defined molecular units. The main feature of this new approach is that each unit is treated as a whole body, which drastically reduces the search space and improves the efficiency, but necessitates the introduction of new variation operators described here. To increase diversity of the population of structures, the initial population and part($\scriptsize{\sim}$20\%) of the new generations are produced using space group symmetry combined with random cell parameters and random positions and orientations of molecular units. We illustrate the efficiency and reliability of this approach by a number of tests (ice, ammonia, carbon dioxide, methane, benzene, glycine and butane-1,4-diammonium dibromide). This approach easily predicts the crystal structure of methane \emph{A} containing 21 methane molecules (105 atoms) per unit cell. We demonstrate that this new approach has also a high potential for the study of complex inorganic crystals as shown on examples of a complex hydrogen storage material Mg(BH$_4$)$_2$ and elemental boron.
\end{abstract}

\maketitle

\section{Introduction}

Structure is the most important piece of information about a material, as it determines most of its physical properties. It was long believed that crystal structures are fundamentally unpredictable \cite{Maddox-1988,Gavezzoti-1994}. However, the situation began to change dramatically in the last decade. As the stable structure corresponds to the global minimum of the free energy, several global optimization algorithms have been devised and used with some success for crystal structure prediction (CSP) – for instance, simulated annealing \cite{SA-1990,SA2-1996}, metadynamics \cite{Metadynamics-2003}, evolutionary algorithms \cite{uspex1,Oganov-ACC}, random sampling \cite{random-sample}, basin hopping \cite{basin-hopping}, minima hopping \cite{minima-hopping}, and data mining \cite{data-mining}. For inorganic crystals, in many cases it is already now possible to predict the stable structure at arbitrary external pressure. Towards the ambition of designing novel materials prior to their synthesis in the laboratory, reliable and efficient prediction of the structure of more complex (in particular, molecular) crystals becomes imperative.

Molecular crystals are extremely interesting because of their applications as pharmaceuticals, pigments, explosives, and metal-organic frameworks \cite{polymorphism2,MOF-2008}. The periodically conducted blind tests of organic crystal structure prediction, organized by Cambridge Crystallographic Data Centre (CCDC) have been the focal point for this community and they reflect steady progress in the field \cite{Blindtest-2000,Blindtest-2002,Blindtest-2005,Blindtest-2009,Blindtest-2011}. The tests show that it is now possible to predict the packing of a small number of rigid molecules, provided there are cheap force fields accurately describing the intermolecular interactions. In these cases, efficiency of search for the global minimum on the energy landscape is not crucial. However, if one has to use expensive \emph{ab initio} total energy calculations or study systems with a large number of degrees of freedom (many molecules, especially if they have conformational flexibility), the number of possible structures becomes astronomically large and efficient search techniques become critically important.

In addition, the nature of weak chemical interactions makes it common that a molecules have a wide variety of ways of packing with lattice energies within a few kJ/molecule of the most stable structure. Thus prediction of such large structures is certainly a challenge, especially if the number of trial structures has to be kept low to enable practical \emph{ab initio} structure predictions. Recent pioneering works \cite{Kim-2009,benzene4,Day:2011}, in particular, using metadynamics \cite{benzene4} offer inspiring examples of this.

Compared to other methods, evolutionary algorithms have special advantages. Exploring the energy surface, such algorithms arrive at the global minimum by a series of intelligently designed moves, involving self-learning and self-improvement of the population of crystal structures \cite{Oganov-ACC}. Our USPEX (Universal Structure Predictor: Evolutionary Xtallography) code \cite{uspex1,uspex2,uspex3,uspex4,uspex5}, proved to be extremely efficient and reliable for atomic crystals, and here we present an extension of this algorithm to complex crystals made of well-defined units. In the following sections, we will mainly discuss molecular crystals. Crystals containing complex ions and clusters can be equally well studied using the methodology developed here, as we show by two tests on challenging systems.

\section{Methodology}

Compared to the prediction of atomic structures, there are several additional considerations to be taken into account for molecular crystals:

i) A typical unit cell contains many more atoms than a usual inorganic structure, which means an explosion of computing costs if all of these atoms are treated independently;

ii) Molecules are bound by weak forces, such as the van der Waals (vdW) interactions, and the inter-molecular distances are typically larger than those in atomic crystals, which leads to the availability of large empty space;

iii) Most of the molecular compounds are thermodynamically less stable than simpler molecular compounds from which they can be obtained (such as H$_2$O, CO$_2$, CH$_4$, NH$_3$). This means that a fully unconstrained global optimization approach in many cases will produce a mixture of these simple molecules, which are of little interest to the organic chemist. To study the packing of the actual molecules of interest, it is necessary to fix the intramolecular connectivity;

iv) Crystal structures tend to be symmetric, and the distribution of structures over symmetry groups is extremely uneven \cite{space-group}. For example, 35\% of inorganic and 45\% of organic materials have the point group 2/\emph{m}. Compared to inorganic crystals, there is a stronger preference of organic crystals to a smaller number of space groups. Over 80\% of organic crystals are found to possess space groups: \emph{P}2$_1$/\emph{c} (36.59\%), \emph{P}-1 (16.92\%), \emph{P}2$_1$2$_1$2$_1$ (11.00\%), \emph{C}2/\emph{c} (6.95\%), \emph{P}2$_1$ (6.35\%) and \emph{Pbca} (4.24\%) \cite{space-group2}.

The first two points indicate that the search space is huge. If we start to search for the global minimum with randomly generated structures, it is very likely that most of the time will be spent on exploring uninteresting disordered structures far away from the global minimum. Fortunately, the last two points suggest a way to improve the efficiency. The point (iii) implies that the true thermodynamic ground state corresponding to most organic compositions is a mixture of simpler molecules, which is of little interest to the organic chemists. The truly interesting problem, packing of the pre-formed molecules, can be solved by \emph{constrained global optimization} – finding the most stable packing of molecules with fixed bond connectivity. This will not only make the global optimization process meaningful, but in the same time will simplify it, leading to a drastic reduction of the number of degrees of freedom and of the search space. Structure prediction (global optimization) must involve relaxation (local optimization) of all structures, and fixing intra-molecular bond connectivity has an added benefit of making structure relaxations cheaper and more robust. Depending on their chemical nature, those molecules shall be treated as fully or partly rigid bodies during the action of evolutionary variation operators and local optimization. Another improvement of the efficiency is achieved by using symmetry in the random generation of new structures - a population of symmetric structures is usually more diverse than a set of fully random (often disordered) structures. Diversity of the population of structures is essential for the success and efficiency of evolutionary simulations.

\begin{figure}
\epsfig{file=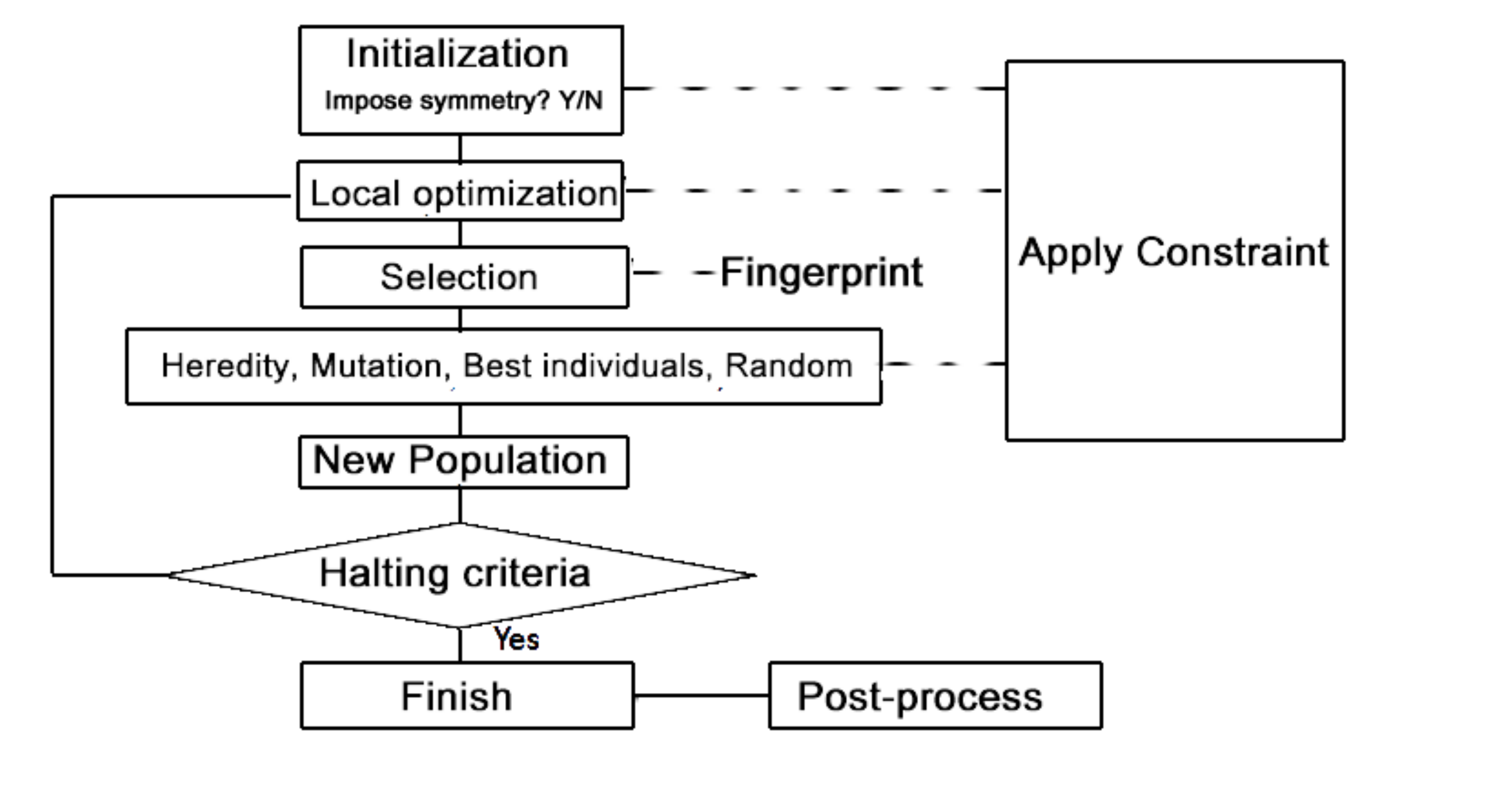, width=0.45\textwidth}
\caption{\label{flowchart} Illustration of the constrained evolutionary algorithm.}
\end{figure}

We have successfully implemented the adapted evolutionary algorithm in the USPEX code. Briefly, our procedure is as follows (as shown in Fig. \ref{flowchart}).

\begin{figure*}
\epsfig{file=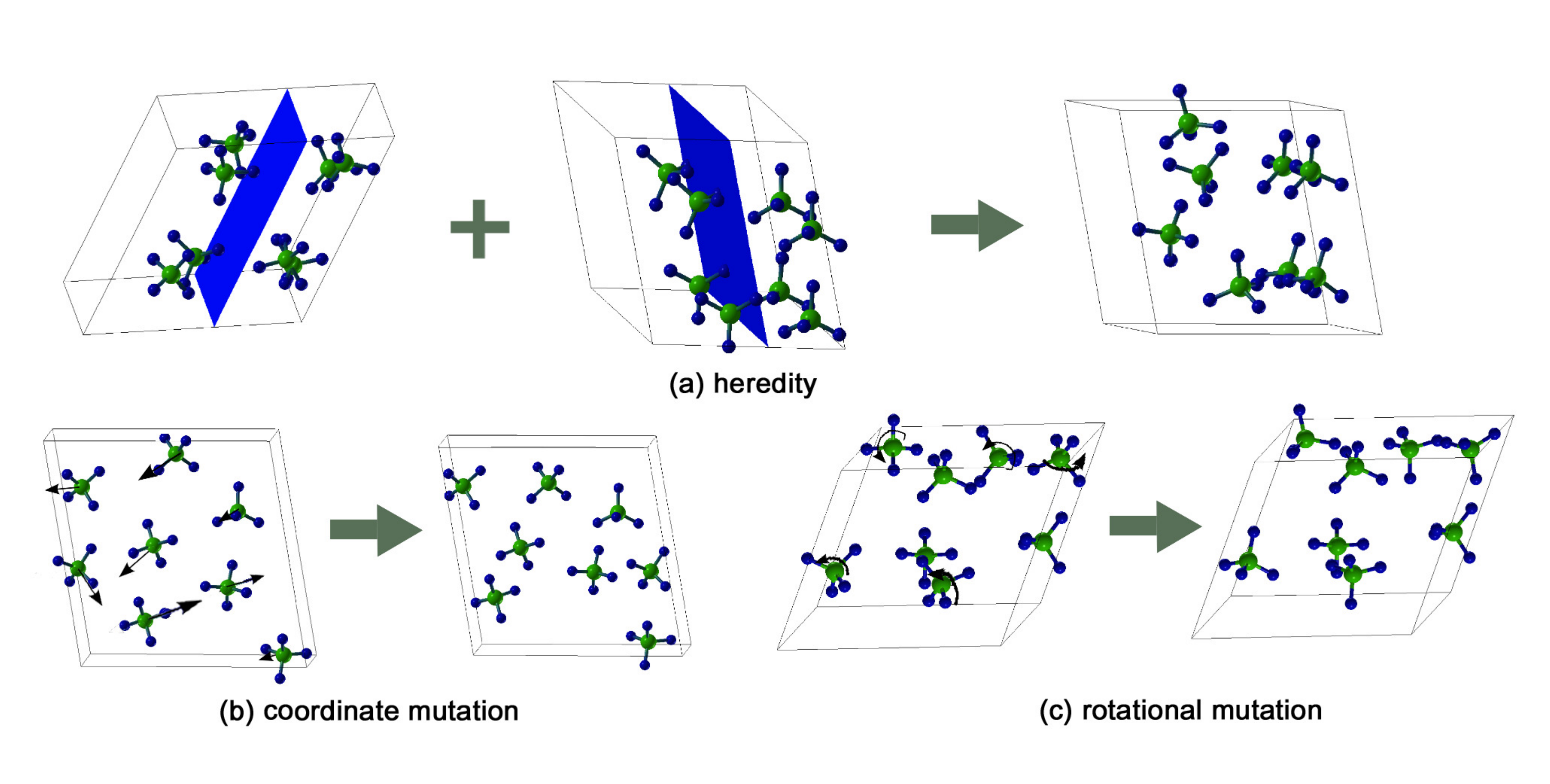, width=0.8\textwidth}
\caption{\label{variation} Illustration of the variation operators: a) heredity; b) coordinate mutation; c) rotational mutation.}
\end{figure*}

1) The initial structures are usually generated randomly, with randomly selected space groups. First, we randomly pick one of 230 space groups, and set up a Bravais cell according to the prespecified initial volume with random cell parameters consistent with the space group. Then one molecule is randomly placed on a general Wyckoff position, and is multiplied by space group operations. If two or more symmetry-related molecules are found close to each other, we merge them into one molecule that sits on a special Wyckoff position and has averaged coordinates of the molecular center and averaged orientational vectors (or random, when averaging gives zero). Adding new molecular sites one by one, until the correct number of molecules is reached, we get a random symmetric structure. During this process, we also make sure that no molecules overlap or sit too close to each other. All produced unit cell shapes are checked and, if necessary, transformed to maximally orthogonal shapes \cite{uspex3}.

2) Structural relaxation is done stepwise from low to high precision. At the initial stages, we employ the SIESTA code \cite{SIESTA} for first-principles simulations, which allows the constrained geometry relaxation. As an option, we can use the DMACRYS code \cite{DMACRYS} for classical calculations. We note that SIESTA provides Z-matrix representation for the molecules\cite{Z-matrix}, enabling the specification of the molecular geometry and its internal degrees of freedom (important when dealing with conformationally flexible molecules). For the final stages of relaxation, we can keep the molecules fully or partly rigid, or allow their complete relaxation (in the latter case, such codes as GULP \cite{gulp} and VASP \cite{vasp} are also supported in USPEX). It is a good strategy to relax the structures in SIESTA with constrained molecular geometry at the beginning stage and then fully relax them using VASP, and here we adopt this strategy for all studied systems. It is well known that DFT within local and semilocal approximations, such as the LDA or GGA, cannot describe vdW dispersion interactions well (e.g., \cite{yanli-2010}), and we therefore used the GGA+D approach that includes a damped dispersion correction \cite{DFT+D}; this approach is known to work well for molecular crystals.

3) At the end of each generation, all structures in the generation are compared using their fingerprints \cite{uspex4} and all non-identical structures are ranked by their (free) energies or (if the calculation is done at T = 0 K, as we do here) enthalpies. There is an important technical aspect: intramolecular contributions are identical for all different packings of the same molecule and thus decrease the discriminatory power of the fingerprint function. Therefore, we remove the intramolecular distances from the computation of the fingerprint function when dealing with molecular crystals.

A certain percentage of higher-energy structures in the population are discarded, and the rest participate in creating the next generation using variation operators detailed below.

To ensure properly constrained global optimization, we not only generate the structures made of the desired molecules, but also check that the bond connectivity has not changed after relaxation - structures with altered connectivity graphs are discarded.

4) Child structures (new generation) are produced from parent structures (old generation) using one of the following variation operators: (i) heredity, (ii) permutation, (iii) coordinate mutation are the same as in atomic crystal structures \cite{uspex1,uspex5}, with the only difference that variation operators act on the geometric centers of the molecules and their orientations, i.e. whole molecules, rather than single atoms, are considered as the minimum building blocks. Since molecules cannot be considered as spherically symmetric point particles, additional variation operators must be introduced: (iv) rotational mutation of the whole molecules, (v) softmutation - a hybrid operator of coordinate and rotational mutation. Fig. \ref{variation} shows how variation operators work in our algorithm. Below we describe how these variation operators were used in our tests.

{\bf Heredity}: This operator cuts planar slices from each individual and combines these to produce a child structure. In heredity, each molecule is represented by its geometric center (Fig. \ref{variation}a) and orientation. From each parent, we cut (parallel to a randomly selected coordinate plane of the unit cell) a slab of random thickness (within bounds of 0.25-0.75 of the cut lattice vector) at a random height in the cell. If the total number of molecules of each type obtained from combining the slabs does not match the desired number of molecules, a corrector step is performed: molecules in excess are removed while molecules in shortage are added; molecules with higher local degree of order have higher probability to be added and lower probability to be removed. This is equivalent to our original implementation of heredity for atomic crystals\cite{uspex1}

{\bf Permutation}: this operator swaps chemical identity in randomly selected pairs of molecules.

{\bf Coordinate mutation}: All the centers of molecules are displaced in random directions, the distance for this movement for molecule \emph{i} being picked from a zero-mean Gaussian distribution with $\sigma$ defined as:

\begin{equation}
\sigma_i = \sigma_{\rm max} \frac{\Pi_{\rm max}-\Pi_i}{\Pi_{\rm max}-\Pi_{\rm min}}
\end{equation}

where $\Pi$ is the local order of the molecule. Thus molecules with higher order are perturbed less than molecules with low order (Fig. \ref{variation}b). We calculate the local order parameter of each molecule from its fingerprint \cite{uspex4}, in the computation of which only centers of all molecules are used. In the tests described here $\sigma_{\rm max}$ represents the order of a typical intermolecular distance.

{\bf Rotational mutation}: A certain number of randomly selected molecules are rotated by random angles (Fig. \ref{variation}c). For rigid molecules, there are only 3 varibles to define the orientation of the molecules. For flexible molecules, we also allow the mutation of torsional angles of the flexible groups. A large rotation can have a marked effect on global optimization, helping the system to jump out of the current local minimum and find optimal orientational ordering.

{\bf Softmutation}: This powerful operator, introduced first for atomic crystals \cite{uspex5}, involves atomic displacements along the softest mode eigenvectors, or a random linear combination of softest eigenvectors. In the context of molecular crystals one must operate with rigid-unit modes and this operator becomes a hybrid operator, combining rotational and coordinate mutations. In this case, the eigenvectors are calculated first, and then projected onto translational and rotational degrees of freedom of each molecule and the resulting changes of molecular positions and orientations are applied preserving rigidity of the fixed intramolecular degrees of freedom. To calculate efficiently the normal modes, we construct the dynamical matrix from bond hardness coefficients \cite{uspex5}. The same structures can be softmutated many times, each time along the eigenvector of a new mode.

At the end of the selection, the best individuals in the last generation (usually up to 5) are kept. To maintain diversity of the population, some fraction (usually 15\% - 30\%) of population is randomly generated with symmetry. While simple random generation does not improve diversity, the use of symmetry does allow a diverse set of structures to be produced.

5) The simulation is stopped once a predefined halting criterion is met. The lowest-energy structures found in USPEX are then carefully relaxed with higher precision using the same level of theory: the all-electron projector-augmented wave (PAW) method \cite{PAW}, as implemented in the VASP code \cite{vasp}, at the level of generalized gradient approximation (GGA) \cite{GGA} for inorganic systems; or dispersion-corrected GGA+D \cite{DFT+D} approximation for organic crystals.  We used the plane wave kinetic energy cutoff of 550 eV and the Brillouin zone was sampled with a resolution of 2$\pi$ $\times$ 0.07 \AA$^{-1}$, which showed excellent convergence of the energy differences, stress tensors and structural parameters.

\section{Tests and applications}

Here we discuss tests of USPEX on system with well-known stable phases and also show how our method finds hitherto unknown structures. The test cases (including ice, methane, ammonia, carbon dioxide, benzene, glycine and butane-1,4-diammonium dibromide, etc) cover a wide range of systems with different molecular shapes (tetrahedral, linear, bent, planar and small biomolecules), and chemical interactions (vdW dispersion, ionic, covalent and metallic bonding, strong/weak hydrogen bonding, $\pi$-$\pi$ stacking, organic and inorganic molecular systems, etc). All the calculations discussed below were performed in the framework of DFT or DFT+D. Driven by the USPEX code, the structures were initially relaxed in SIESTA with constrained molecular geometry and then fully relaxed in VASP at the final stages.

\subsection{Ice}
Ice (H$_2$O) is an archetypal hydrogen-bonded molecular crystal. The orientations of hydrogen bonds locally obey the well-known ice rules, that is, each oxygen atom is tetrahedrally bonded to four hydrogen atoms, by two strong covalent intra-molecular bonds and two much weaker inter-molecular bonds (hydrogen bonds). Given the enormous number of possibilities of placing and orienting (even under ice rules) water molecules, the prediction of the ice structure is a complex task: according to Maddox(1988), it is \emph{still thought to lie beyond the mortals' ken}.

The normal crystalline form of ice, ice I$_{\emph h}$, is disordered and has hexagonal symmetry, with oxygen atoms arranged in a hexagonal diamond motif (a cubic diamond-type ice I$_{\emph c}$ is also known experimentally) with randomly oriented hydrogen bonds. In experiment \cite{ice1,ice2}, ice XI (the ordered version of ice I$_{\emph h}$), found to be the most stable polymorph at 1 atm and low temperatures, but the transformation from disordered ice I$_{\emph h}$ to ordered ice XI is kinetically hindered, and this is why special approaches are needed for experimental preparation of ice XI \cite{ice1}.

\begin{figure*}
\epsfig{file=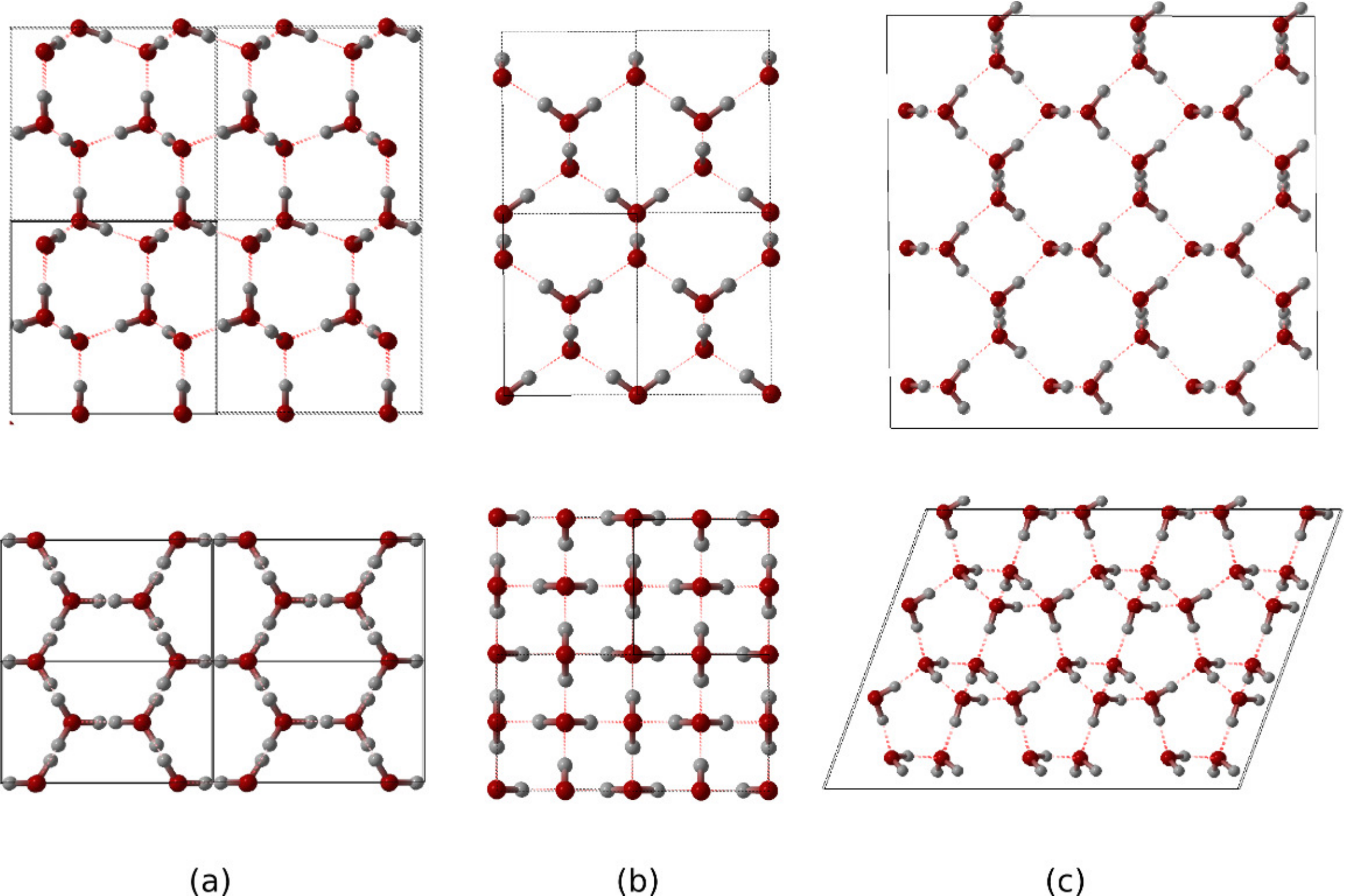, width=0.6\textwidth}
\caption{\label{ice} Ice polymorphs at 1 atm found by USPEX. a) Ice XI, derived from ice I$_{\emph h}$, space group \emph{Cmc}2$_1$, a=4.338 \AA, b=7.554 \AA, c=7.094 \AA, O1(0,0.6651,0.0623), O2(0.5,0.8328,-0.0622), H1(0,0.6638,0.2045), H2(0,0.5373,-0.0191), H3(0.6865,-0.2329,-0.0140); b) tetragonal phase, derived from ice I$_{\emph c}$, space group \emph{I}4$_1$\emph{md}, a=4.415 \AA, c=6.008 \AA, O(0,0.5,0.0006), H(0.1839,0.5,0.1000); c) bct-4 like ice, space group \emph{Cm}, a=4.472 \AA, b=10.451 \AA, c=5.744 \AA, $\beta$=111.3$^\circ$, O1(0,0,0), O2(0.3691,0,0.7197), O3(0.6647,0.3192,0.3605), H1(0.7683,0,0.8871), H2(0.1317,0,0,8908), H3(0.4705, 0.2691, 0.3567), H4(0.0923,0.8787,0.2133), H5(0.3018,0.0751,0.6030).}
\end{figure*}

\begin{figure}
\epsfig{file=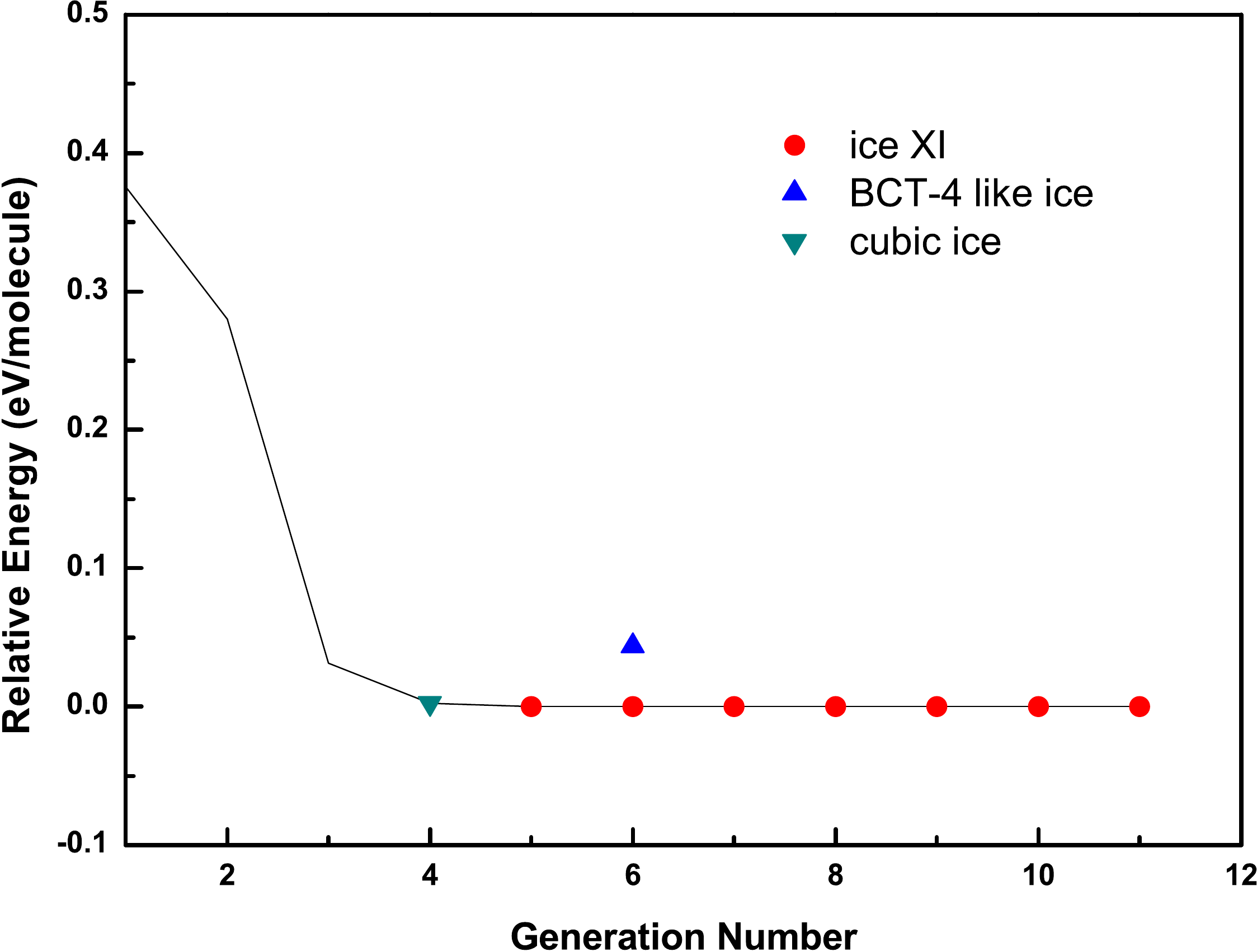, width=0.4\textwidth}
\caption{\label{ice2} Prediction of the crystal structure of ice at 0 GPa.  The lowest energy at each generation is shown relative to the ground state. Each generation contains 30 structures. The ground state structure ice XI was found at 4th generation. We also found cubic ice and bct4-like ice at the same calculation.}
\end{figure}

With variable-cell USPEX simulations for a 4-molecules cell at 1 atm, we indeed identify ice XI as the most stable polymorph (Fig. \ref{ice}a). This structure was found within just 4 generations, after relaxing $\scriptsize{\sim}$160 structures. Fig. \ref{ice2} shows how the lowest energy changed from generation to generation in our calculation. This purely quantum-mechanical calculation required less than 1 day on 8 cores of a Dell XPS desktop PC. Apart from ice XI, we found several remarkable structures in the same run.

An ordered version of ice I$_{\emph c}$ \cite{ice3}, a tetragonal phase with a cubic diamond-type oxygen sublattice (Fig. \ref{ice}b), was found to be energetically competitive with ice XI. At both the GGA and GGA+D levels of theory, its energy is only 2 meV per molecule above that of ice XI. We have also found an interesting low-energy metastable polymorph (Fig. \ref{ice}c), where the oxygen sublattice has topology of the hypothetical bct4 allotrope of carbon \cite{ice4,ice5}. The bct4-like structure of ice was also found from molecular dynamics simulation of the water's adsorption on the surface of hydroxylated $\beta$-cristobalite \cite{ice6}. Proton ordering lowers its symmetry from \emph{I}4/\emph{mmm} to \emph{Cm}.

\subsection{Methane}
Methane, the simplest of saturated hydrocarbons, is an important constituent of gas-giant planets Uranus and Neptune \cite{methane1}. The high-pressure behavior of methane is of extreme importance for fundamental chemistry, as well as for understanding the physics and chemistry of planetary interiors.

The tetrahedral CH$_4$ molecules are interacting practically only by vdW dispersion forces with each other. In spite of the simplicity of the molecule, the phase diagram of methane is still not well established \cite{methane2,methane3,Nakahata-1999-CPL,Sun-2009-CPL}. Different experiments on methane were conducted during the last few decades, resulting in a complex phase diagram drawn by Bini and Pratesi \cite{methane2}. Of the nine solid phases in the diagram, only the structures of phases I, II, and III have been determined, while phases II, III, IV, V, and VI, only exist below 150 K and at moderate pressures. CH$_4$ is expected to become chemically unstable and decompose at megabar pressures \cite{methane5}.

The high-pressure phases of solid methane above 5 GPa have been the subject of numerous experimental and theoretical studies, however, the understanding are still incomplete. Bini and Pratesi \cite{methane2} based on IR and Raman data, proposed a tetragonal crystal structure for the phase \emph{A}, while high-pressure X-ray powder diffraction experiments suggested that the unit cell contains 21 molecules in the pseudocubic rhombohedral unit cell \cite{Nakahata-1999-CPL}.

\begin{figure*}
\epsfig{file=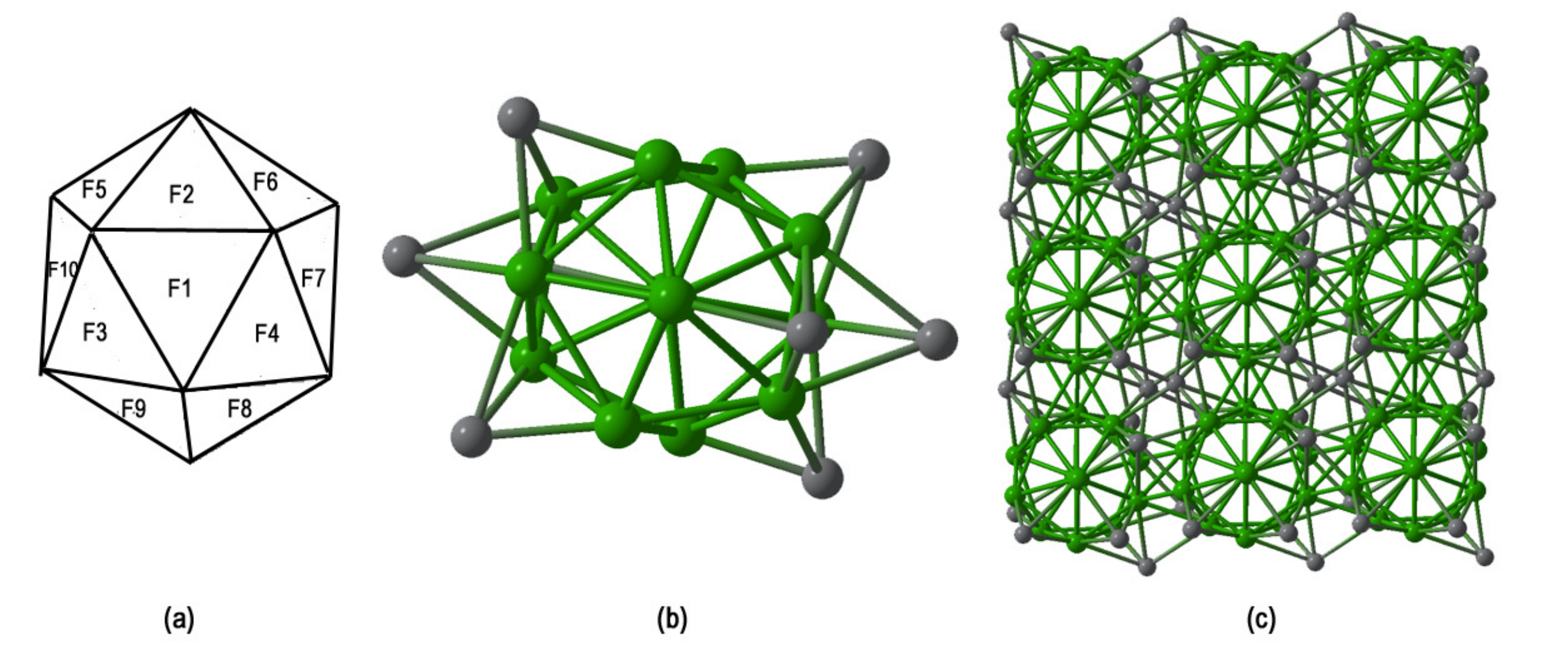, width=0.8\textwidth}
\caption{\label{21-mol-structure}Structures of methane: (a) illustration of possible sites around the icosahedron, (b)21-molecule rhombohedral methane, with F1, F2, F3, F4 sites occupied; (c) view of the icosahedral packing in the rhombohedral methane (space group: R-3); Two C sublattices are marked by different colors in 21-molecule cell (outside icosahedral, green; within icosahedral, grey)}
\end{figure*}

We performed structure prediction simulation for CH$_4$ using experimental cell parameters \cite{Sun-2009-CPL} (a=8.36 {\AA} at 11 GPa). We indeed found the best structure to possess a rhombohedral symmetry, and this structure was found within 8 generations, and is characterized by the icosahedral packing of methane molecules, and this packing fully explains the unusual number \emph{21} of molecules in the cell: \emph{1} molecule is located in the center of the unit cell, \emph{12} molecules around it form an icosahedron, and the remaining \emph{8} molecules are located above the triangular faces of the icosahedron (Fig. \ref{21-mol-structure}). A rhombohedral model, very similar to ours, was recently proposed, on the basis of neutron scattering experiments \cite{methane3}: the only difference is that our model has orientationally disordered molecules (as is also most likely to be the case in reality: furthermore, this model has a lower energy), while Ref. \cite{methane3} assumed orientationally ordered molecules. The essential icosahedral character of the structure was not mentioned by Ref. \cite{methane3}, but can be clearly seen on close inspection of their results.

\subsection{Ammonia}
Bonding in NH$_3$ is intermediate between hydrogen bonded tetrahedral structure of H$_2$O, and vdW-bonded close-packed structure of CH$_4$. Weak hydrogen bonding between neighboring ammonia molecules results in a pseudo-close-packed arrangement in the solid state \cite{ammonia1}. It is extremely interesting to understand the nature of hydrogen bonding in crystalline ammonia, and the properties of ammonia under pressure are of fundamental interest, as compressed ammonia has a significant role in planetary physics{\cite{methane1}}.

Bonding in NH$_3$ is intermediate between hydrogen bonded tetrahedral structure of H$_2$O, and vdW-bonded close-packed structure of CH$_4$. Weak hydrogen bonding between neighboring ammonia molecules results in a pseudo-close-packed arrangement in the solid state \cite{ammonia1}. It is extremely interesting to understand the nature of hydrogen bonding in crystalline ammonia; properties of ammonia under pressure are of fundamental interest, as compressed ammonia has a significant role in planetary physics{\cite{methane1}}.

At room temperature, ammonia crystallizes at 1 GPa in a rotationally disordered, face-centered-cubic phase (phase III) \cite{ammonia1,ammonia3,ammonia4}. X-ray and neutron studies have yielded information about the equation of state and structures of solid ammonia. The low-\emph{P, T} phase I of ammonia undergoes a first-order phase transition into phase IV at about 3 - 4 GPa and then into phase V at about 14 GPa. Phase I has a cubic structure with space group \emph{P}2$_1$3, while phase IV has been identified as the orthorhombic structure with space group \emph{P}2$_1$2$_1$2$_1$. Phase V might have the same space group as phase IV (an isosymmetric phase transition).

\begin{figure*}
\epsfig{file=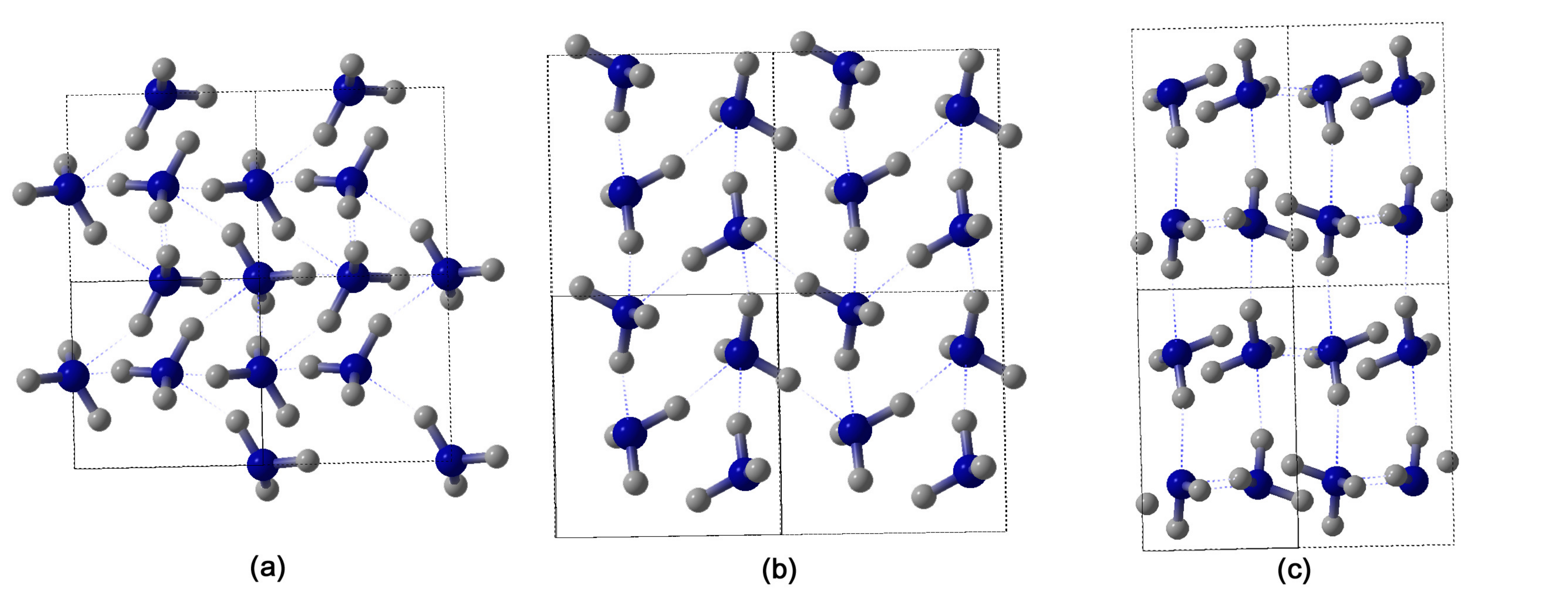, width=0.75\textwidth}
\caption{\label{nh3} Crystal structures of ammonia. a) \emph{P}2$_1$3 phase (stable at 1 - 6 GPa, \emph{Z}=4); b) \emph{P}2$_1$/c phase (stable at 6 - 8.5 GPa, \emph{Z}=4); c) \emph{P}2$_1$2$_1$2$_1$ phase (stable at 8.5 - 60 GPa, \emph{Z}=4)}
\end{figure*}

We carried out variable-cell structure prediction calculations at 5, 10, 25, 50 GPa. At low pressures (5 GPa), we found that the \emph{P}2$_1$3 structure to be stable (Fig. \ref{nh3}a), in good agreement with the experiment. At high pressures, USPEX without applying symmetry in the initialization still easily found \emph{P}2$_1$3 structure, however, failed to get the ground state structure \emph{P}2$_1$2$_1$2$_1$ phase in a simulation with up to 20 generations. The energies of whole-molecule rotation are very small compared to intra-molecular bonding energies, thus making the process of finding correct molecular orientations extremely difficult. This indicates that the energy landscape of ammonia is actually very flat. To enhance the searching efficiency, we initialized the first generation using random symmetric structures. And to retain diversity of the population, 30\% of each new generation was produced by random symmetric mechanism. In this case, the ground state structure (Fig. \ref{nh3}c) appeared within 6 generations ($\scriptsize{\sim}$210 structural relaxations). In addition, we also found the \emph{P}2$_1$/\emph{c} phase (Fig. \ref{nh3}b) reported before \cite{ammonia2}.

\subsection{Carbon dioxide}
The CO$_2$ molecule has a special significance because it is very abundant in nature and is a model system involving $\pi$ bonding and \emph{sp}-hybridization of carbon atoms. Similar to methane, carbon dioxide is a vdW crystal with strong (weak) intra-molecular (inter-molecular) interactions at low pressures \cite{co1}. At room temperature and 1.5 GPa, CO$_2$ crystallizes as dry ice, with a cubic \emph{Pa}3 structure. At pressures between 12 and 20 GPa, CO$_2$-I transforms to the orthorhombic CO$_2$-III \cite{co2,co3,co4}. According to the theoretical calculation, CO2-(III) is metastable, while CO2-(II) with the \emph{P}4$_2$/\emph{mnm} symmetry is believed to be thermodynamically stable \cite{co7} It is known that above 20 GPa a non-molecular phase (called phase V) with tetrahedrally coordinated carbon atoms becomes stable \cite{co1}.

\begin{figure}
\epsfig{file=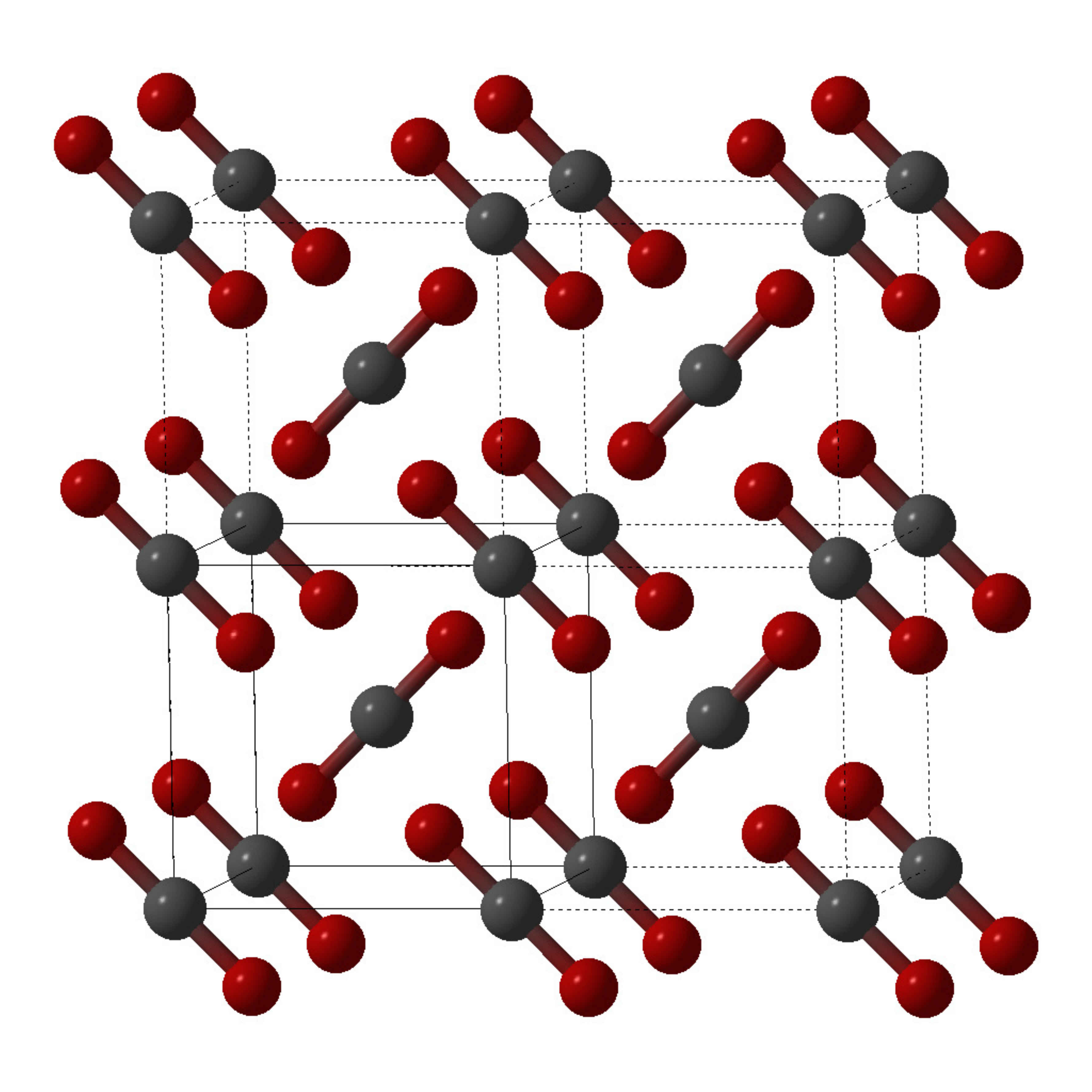, width=0.45\textwidth}
\caption{\label{co2} Crystal structure of CO$_2$ II (space group: \emph{P}4$_2$/\emph{mnm}, \emph{Z}=2). }
\end{figure}

In the previous prediction \cite{co5}, unconstrained USPEX calculations succeeded in finding the correct CO$_2$ structures in a wide pressure range. By applying molecular constraint, we have found the \emph{P}4$_2$/\emph{mnm} phase (Fig. \ref{co2}) quicker, just in two generations ($\scriptsize{\sim}$80 structural relaxations). \emph{P}4$_2$/\emph{mnm} phase remains the most stable structure made of discrete CO$_2$ molecules at least up to 80 GPa.  Both experiment \cite{co6} and theory \cite{co5,co7} show that CO$_2$ polymerizes above 20 GPa, while the molecular form (\emph{P}4$_2$/\emph{mnm} phase) exists as a metastable form at low temperatures and higher pressures. This examples shows how imposing constraints gives the most stable molecular form, while unconstrained search finds the global minimum (which for CO$_2$ is non-molecular above 20 GPa). Both cases correspond to situations that are experimentally achievable, and thus important.

\subsection{Benzene}
Benzene is the simplest aromatic compound, and it has a purely planar molecule, the packing of which is stabilized by $\pi$-$\pi$ interactions. The crystal structure of benzene is one of the most basic and most actively investigated structures. The first proposed phase diagram was very complex and contained six solid phases \cite{benzene1}. However, recent experimental studies simplified it \cite{benzene2,benzene3}. At normal conditions, benzene crystallizes in the orthorhombic phase I (\emph{Pbca}). A monoclinic phase II (\emph{P}2$_1$/\emph{c}), with two molecules per unit cell was idenfied above 1.75 GPa. Phase II is stable up to the onset of a chemical reactions (at 41 GPa and 298 K).

\begin{figure*}
\epsfig{file=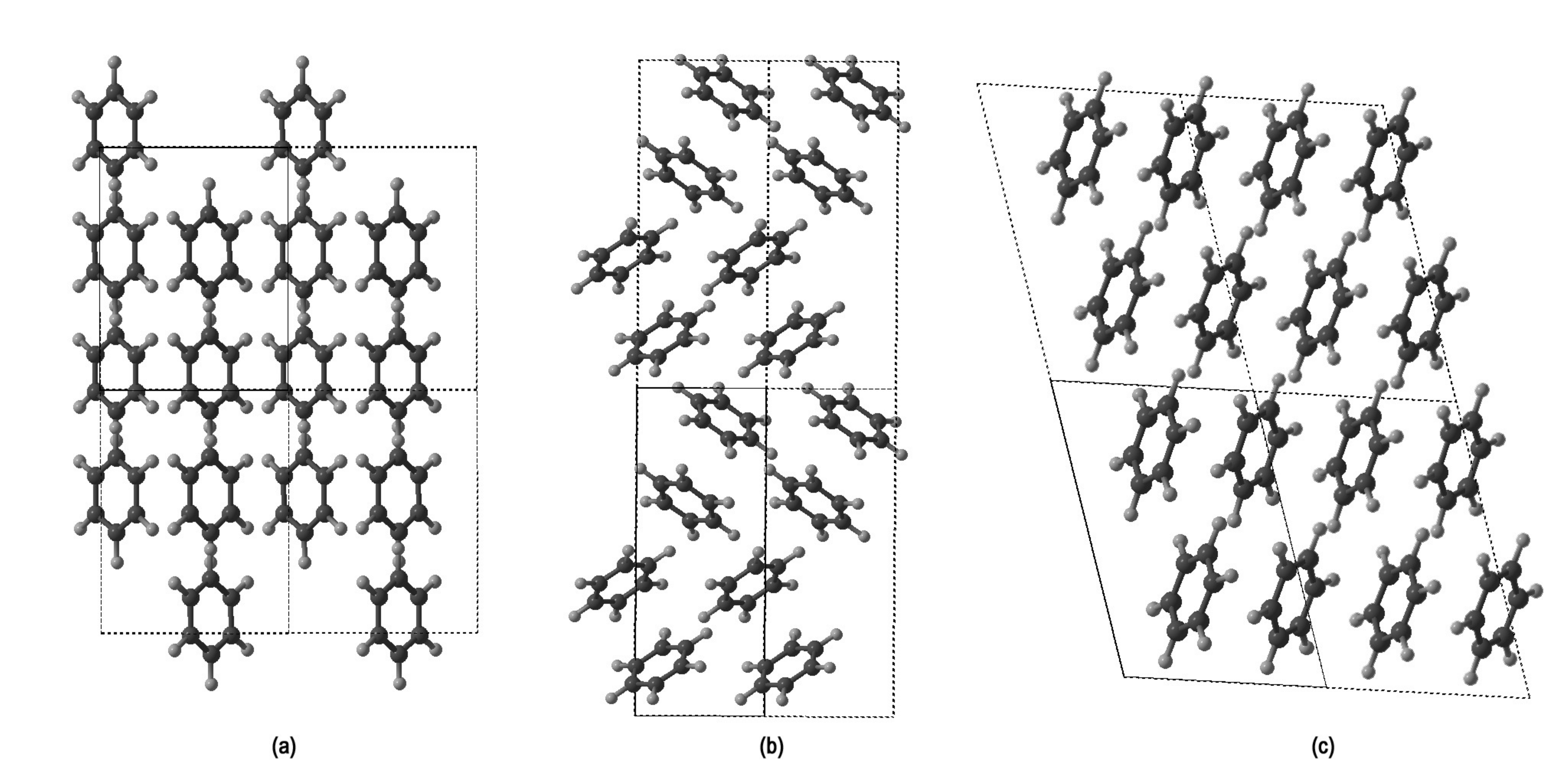, width=0.95\textwidth}
\caption{\label{benzene} Crystal structures of benzene (a)orthorhombic phase I (\emph{Pbca}, \emph{Z}=4); (b)tetragonal phase II (\emph{P}4$_3$2$_1$2, \emph{Z}=4); (c) monoclinic phase (\emph{P}2$_1$/c, \emph{Z}=2).}
\end{figure*}

In our simulation we started with the same emperical potential \cite{GROMOS96} as used in a recent metadynamics study \cite{benzene4}, and we reproduced the multiple phases of benzene found there and corresponding to the old phase diagram. This potential was calibrated at normal conditions and may fail at high pressure. Its predicted many stable phases at different pressures (this is consistent with the old phase diagram, but most of these phases should be metastable according to the new experiment). To remedy this, we repeated our structure prediction runs at the level of DFT+D. We performed the calculation at 0, 5, 10, 25 GPa with \emph{Z}=4. In our simulation, the experimentally observed orthorombic phase (\emph{Pbca}) was identified as the most stable phase at 0 GPa, and then it transforms to \emph{P}4$_3$2$_1$2 phase at 4 GPa. We also found the monoclinic phase (\emph{P}2$_1$/\emph{c}) (Fig. \ref{benzene}) as the ground state above 7 GPa. Our DFT+D results give fewer stable phases, in agreement with the new phase diagram - the only difference is in the \emph{P}4$_3$2$_1$2 phase. This phase is experimentally known, and according to the latest experimental results \cite{benzene2,benzene3} is metastable. Previous DFT calculation \cite{benzene5} suggested this phase to be stable at pressures 4-7 GPa, which is consistent with our results. Thermodynamic stability of the \emph{P}4$_3$2$_1$2 phase needs to be revisited experimentally.

\subsection{Glycine}
Glycine, with the formula NH$_2$CH$_2$COOH, is the smallest of 20 aminoacids commonly found in proteins. Aminoacids are important in nutrition and widely used in the pharmaceutical industry.

\begin{figure*}
\epsfig{file=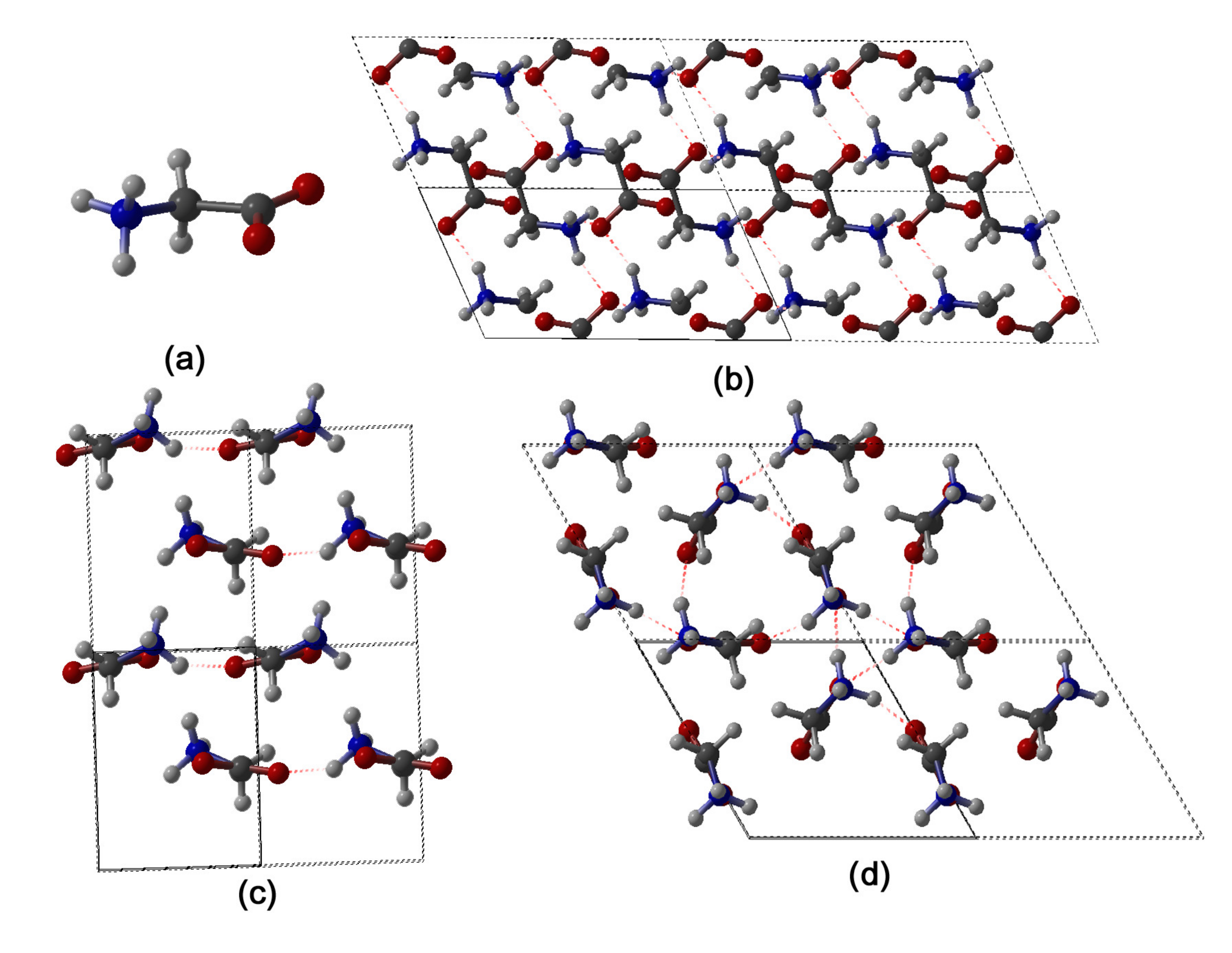, width=0.75\textwidth}
\caption{\label{glycine} Glycine polymorphs found by USPEX. a) representation of glycine zwitterion; b) $\alpha$-glycine at 2 GPa(\emph{Z}=4, a=5.390 \AA, b=5.911 \AA, c=10.189 \AA, $\beta$=113.2 $^\circ$); c) $\beta$-glycine at 0.4 GPa(\emph{Z}=2, a=5.372 \AA, b=6.180 \AA, c=5.143 \AA, $\beta$=111.9 $^\circ$); d) $\gamma$-glycine at 1 GPa(\emph{Z}=3, a=b=7.070 \AA, c=5.490 \AA).}
\end{figure*}

The polymorphism of glycine was intensely studied \cite{glycine1,glycine2,glycine3,glycine4,glycine5,glycine6,glycine7}. Glycine is known to crystallize in four polymorphs with space groups \emph{P}2$_1$/\emph{c}, \emph{P}2$_1$, \emph{P}3$_2$ and \emph{P}2$_1$/\emph{c} which are labeled $\alpha$, $\beta$, $\gamma$ and $\sigma$, respectively \cite {glycine1}. The $\alpha$, $\beta$, and $\gamma$ phases are found at ambient pressure, with $\alpha$ and $\beta$ phases being metastable with respect to the $\gamma$ phase. $\sigma$ glycine has recently been found to form under pressure \cite{glycine3}. In the gas phase, glycine is in a nonionic form, while in all four of the crystal structures glycine is zwitterionic (as shown in Fig. \ref{glycine}a). In this form, an -NH$_3$$^+$ group on one ion electrostatically interacts with a -COO$^-$ group on a neighboring ion. Although zwitterionization causes an increase in energy with respect to the gas-phase molecule, it is thought that the zwitterionic crystals are stabilized by the increase in number of hydrogen bonds that can be formed in comparison to the number that would be formed in the nonionic case.

Since the glycine zwitterion only has the point symmetry \emph{C}$_1$ (i.e. no symmetry), structure prediction of glycine is more challenging compared with benzene. We performed variable cell prediction at 1 GPa with 2 - 4 molecules per cell. Without any experimental information, we found $\beta$-glycine (Fig. \ref{glycine}c) as a metastable structure with \emph{Z} = 2; and $\gamma$-glycine (Fig. \ref{glycine}d) as the best structure with \emph{Z} = 3. We also found $\alpha$-glycine as a metastable form in the calculation with \emph{Z} = 4 (Fig. \ref{glycine}b) at 2.0 GPa. This shows the power of our evolutionary search method. However, GGA+D results show that $\alpha$ glycine possesses the lowest enthalpy, while $\gamma$ and $\beta$ phase are 20 meV/molecule and 30 meV/molecule higher, respectively. Yet, the experimental results demonstrated the relative thermodynamic stability to be $\gamma$ $>$ $\alpha$ $>$ $\beta$. This shows the need for better ways of computing intermolecular interaction energies.

\subsection{Butane-1,4-diammonium dibromide}
The molecules we discussed so far are rigid or nearly rigid. Is it possible to use this approach to study the packing of flexible molecules? To investigate this, we applied it to the prediction of crystal structure of butane-1,4-diammonium dibromide, in which Br$^{-}$ and {C$_4$H$_{14}$N$_2$}$^{2+}$ can be described as two molecular units that form the structure.

By using the experimental cell parameters \cite{Blerk-2006}, we indeed observed numerous structures with different conformations of the {C$_4$H$_{14}$N$_2$}$^{2+}$ molecular ion. USPEX firstly found the energetically favorable conformation, and then identified the ground state structure at the 12th generation (about 500 structural relaxations): \emph{P}2$_1$/c butane-1,4-diammonium dibromide. In this structure, as shown in Fig. \ref{butane-1,4}, the organic hydrocarbon chains are found to pack in a herringbone-type stacking with hydrogen bonds to the Br$^{-}$. Each Br$^{-}$ anion is surrounded by four --NH$_{3}$$^+$ groups. During the process of rotational mutation, both the orientation of the whole molecular group and its flexible torsional angles are allowed to change. A large fraction of rotation ($\scriptsize{\sim}$ 40\%) of the molecules is found to greatly speed up the prediction. This success confirmed that our constrained evolutionary algorithm can be straightforwardly adapted to deal with flexible molecules.

\begin{figure*}
\epsfig{file=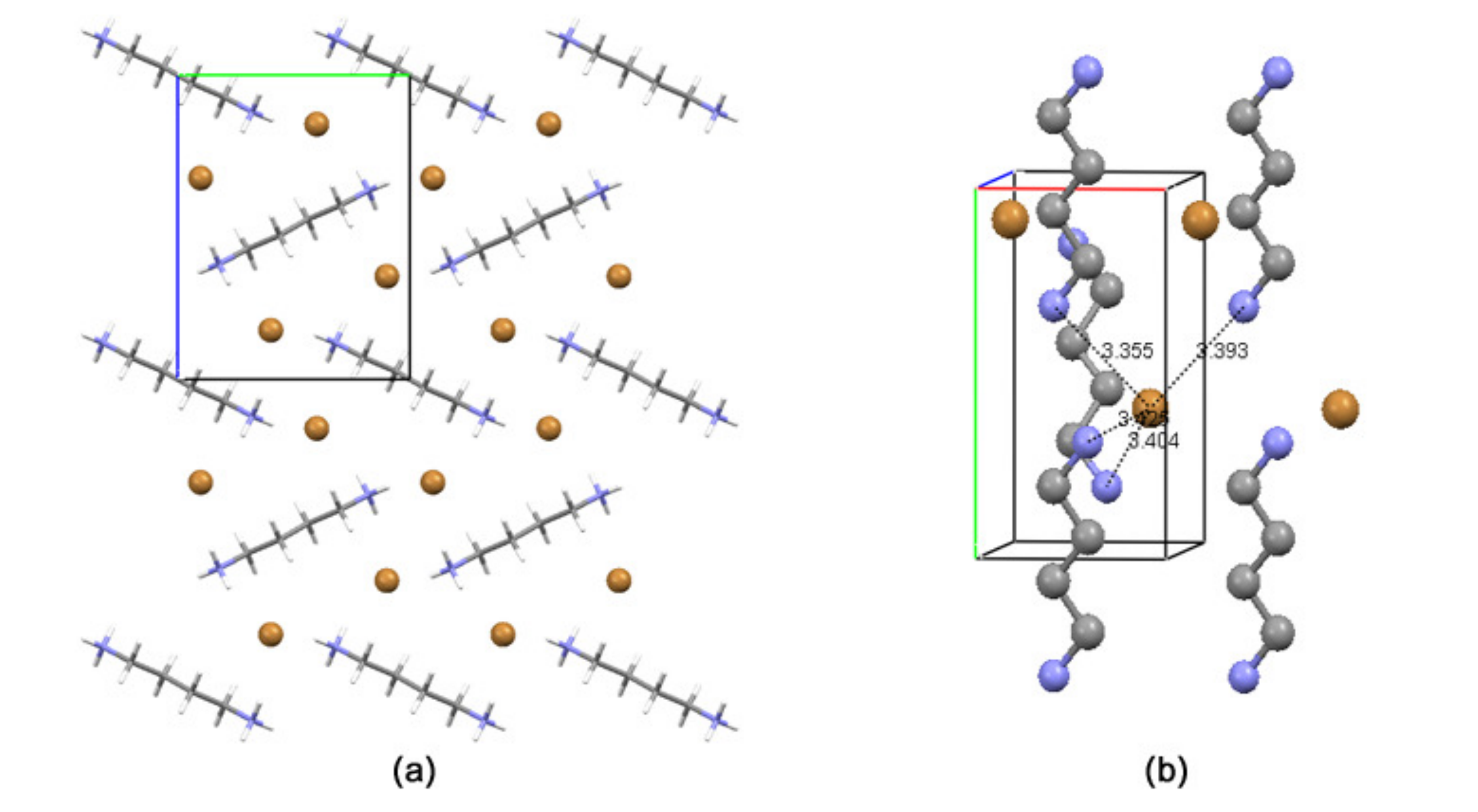, width=0.75\textwidth}
\caption{\label{butane-1,4} Butane-1,4-diammonium dibromide polymorph found by USPEX (space group: \emph{P}2$_1$/c, \emph{Z}=2). a), representation of network, view from \emph{a} axis; b) Br$^-$ coordination environment, view from \emph{b} axis. For clarity, hydrogen atoms are not shown in b. The {C$_4$H$_{14}$N$_2$}$^{2+}$ molecular ion has 6 flexible angles, and the unit cell of the stable polymorph contains 44 atoms.}
\end{figure*}

\subsection{Inorganic crystals}
Apart from molecular crystals, this new approach is also applicable to inorganic crystals with complex ions or clusters. Below are a few illustrations.

\subsubsection{Complex ionic solids: example of hydrogen storage materials}
Reversible hydrogen storage materials recently attracted great interest \cite{Hydrogen-store:2001}. Two groups of complex metal hydrides: alumohydrides containing AlH$_4$ groups and borohydrides with BH$_4$ groups have been recently under intensive study \cite{MgBH1,MgBH2,MgBH3,MgBH4}. Numerous dehydriding and rehydriding processes have been predicted theoretically and tested experimentally. In a good candidate material, dehydridation should happen at acceptably low temperatures. Structure prediction for such systems can guide the experimentalists to synthesize the desired compounds in the laboratory.

The crystal structure of Mg(BH$_4$)$_2$ has been extensively investigated. It was experimentally solved, and found to be extremely complex (330 atoms per unit cell for the low-temperature phase with \emph{P}6$_1$ symmetry \cite{MgBH1}). Recent theoretical work then predicted a new body-centered tetragonal phase (with \emph{I}-4\emph{m}2 symmetry), which has slightly lower energy than \emph{P}6$_1$ phase; it was found using the prototype electrostatic ground-state approach (PEGS) \cite{MgBH2}. Later, based on the prototype structure of Zr(BH$_4$)$_4$, another orthorhombic phase with \emph{F}222 symmetry was found to have even lower energy than all previously proposed structures \cite{MgBH4}.

\begin{figure*}
\epsfig{file=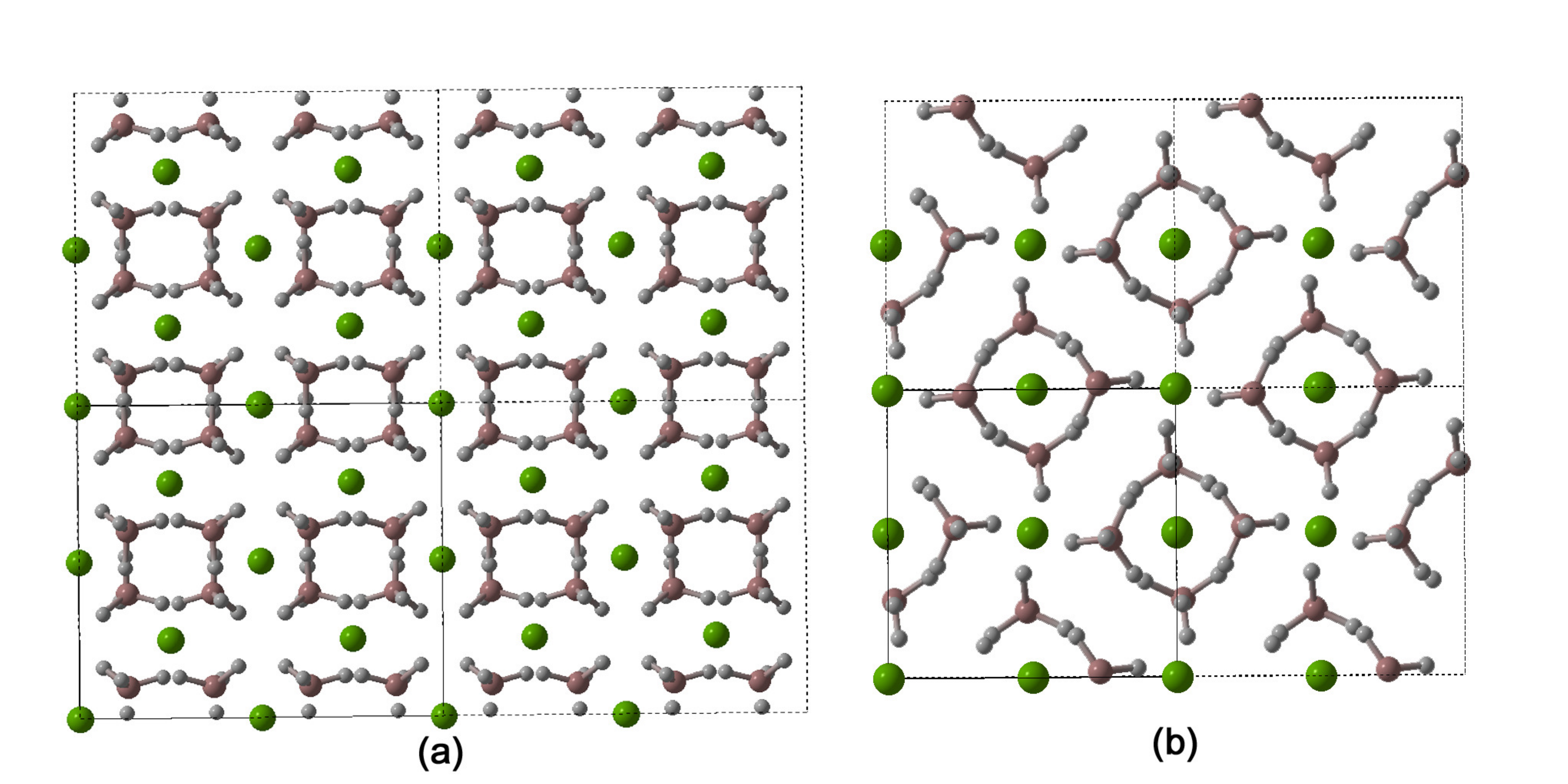, width=0.8\textwidth}
\caption{\label{MgBH4} Mg(BH$_4$)$_2$ polymorphs found by USPEX. a) \emph{F}222 phase; b) \emph{I}4$_1$22 phase. }
\end{figure*}

In general, the previous theoretical discoveries of novel Mg(BH$_4$)$_2$ phases were conducted either by \emph{ad hoc} extensive searching or by chemical intuition. However, our evolutionary algorithm does not rely on any prior knowledge except chemical composition, and could be particularly useful for predicting stable crystal structures for these complex metal hydride systems. If we consider the BH$_4$$^-$ ion as a molecular group, the search space would be dramatically reduced. Within 10 generations ($\scriptsize{\sim}$400 structure relaxations), USPEX found the \emph{F}222 phase (Fig. \ref{MgBH4}a) as the most stable structure at ambient pressure. In addition, \emph{I}-4\emph{m}2 (Fig. \ref{MgBH4}b) was also found by USPEX in the same calculation, with enthalpy less than 1.2 meV/atom above that of \emph{F}222 phase. Compared to the previous work, our method is clearly more universal and robust, enables efficient structure prediction for complex molecular systems, both organic and inorganic.

\subsubsection{Cluster-based crystals: example of elemental boron}
\begin{figure*}
\epsfig{file=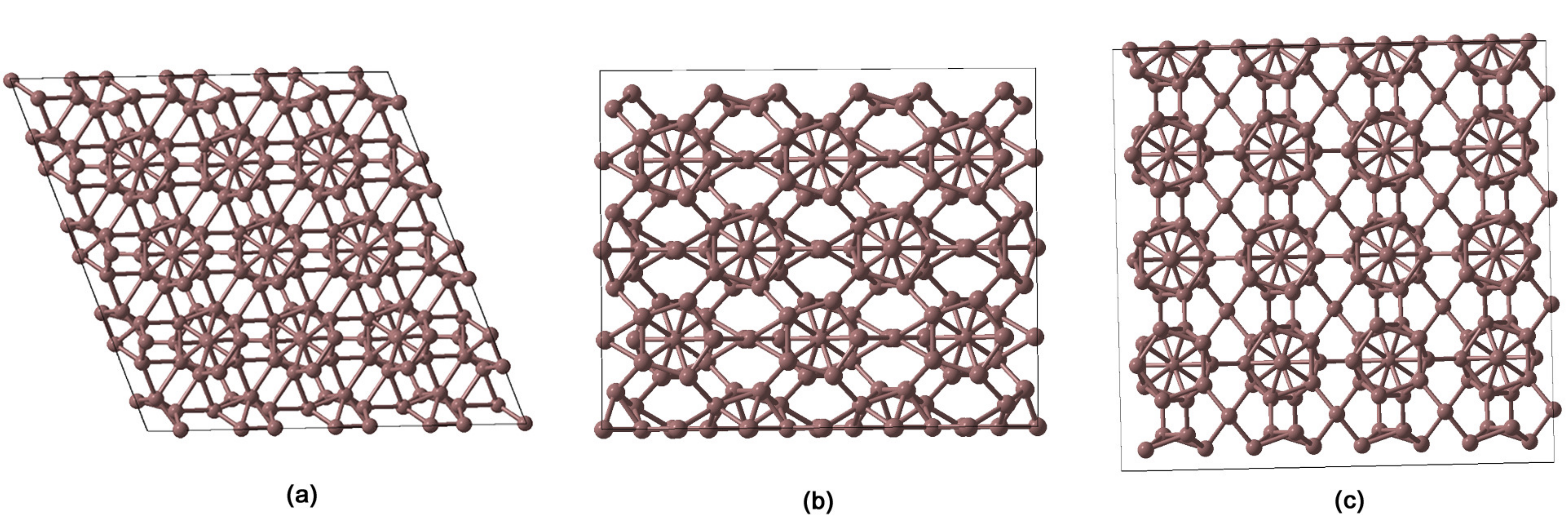, width=1.0\textwidth}
\caption{\label{boron} Crystal structures of boron. a) $\alpha$-B$_{12}$; b) $\gamma$-B$_{28}$; c) novel metastable B$_{52}$ phase, space group \emph{Pnn2}, a=8.868 \AA, b=8.777 \AA, c=5.000 \AA. B1(0.5777,0.7728,0), B2(0.9187,0.7321,0.3222), B3(0.7464,0.7488,0.4978), B4(0.5902,0.6690,0.3087), B5(0.6243,0.8678,0.3055), B6(0.8209,0.9090,0.3202), B7(0.8698,0.6288,0.0229), B8(0.7835,0.5798,0.3273), B9(0.6684,0.5795,0.0182), B10(0.7190,0.9189,0.0056), B11(0.8991,0.8309,0.0115), B12(0.7461,0.7461,0.8302), B13(0,0,0.7529), B14(0,0.5,0.9161)}
\end{figure*}

Boron, located in a unique position of Periodic Table, is an element of chemical complexity due to the subtle balance between localized and delocalized electronic states. All known structures of boron contain icosahedral B$_{12}$ clusters. Recent experiment \cite{Boron:2009} found a new phase of pure boron ($\gamma$-B$_{28}$) at pressures above 10-12 GPa, and its structure was solved using USPEX with fixed experimental cell dimensions \cite{Boron:2009}. Surprisingly, $\gamma$-B$_{28}$ showed different chemistry compared with all the other elemental boron allotropes. In the $\gamma$-B$_{28}$ structure (Fig. \ref{boron}b), the centers of the B$_{12}$ icosahedra form a distorted cubic close packing as in $\alpha$-$B_{12}$ (Fig. \ref{boron}a); but with all octahedral voids are occupied by $B_2$ pairs. The $\gamma$-B$_{28}$ structure resembles a NaCl-type structure, with the B$_{12}$ icosahedra and B$_2$ pairs as `anions' and `cations'. Finding this structure without fixing cell parameters was reported to be exceedingly difficult \cite{Ho:2010}, but latest methodological developments enable this (Lyakhov \emph{et al.}, unpublished). However, the problem can be made very simple if we recall that all boron phases contain B$_{12}$ icosahedra. Here we treated B$_{12}$ icosahedral and B$_2$ pairs as separate rigid units, and performed structure prediction runs at different numbers of B$_{12}$ and B$_2$ units (2:1, 1:1, 2:2, 2:4, etc) at ambient conditions. We could easily find $\gamma$-B$_{28}$ within 2 - 3 generations or $\scriptsize{\sim}$100 structural relaxations. Meanwhile, we observed a set of low-energy and chemically interesting structures with different proportions of B$_{12}$ and B$_2$. For instance, the novel metallic phase B$_{52}$ with the \emph{Pnn2} symmetry (Fig. \ref{boron}c) was calculated to be only 12 meV/atom higher in energy than $\gamma$-B$_{28}$ at atmosphere pressure. Its energy is lower than those of the experimentally observed phases (such as the T-50 phase \cite{Hoard-JACS-1958}) and this example shows that our method can be used for even non-molecular and inorganic solids that contain clusters or complex ions.

\section{Discussion and conclusions}

In the original version of USPEX \cite{uspex1}, the stable crystal structure was assembled from individual atoms, which was also shown to work well for atomic crystals and also for simple molecular systems (like carbon dioxide, water, urea). However, it is clear that for molecular crystals improvements of the efficiency can be made if the structure is assembled from whole molecules rather than individual atoms. This is confirmed by the present study. Our \emph{constrained global optimization} method allows one to find the stable crystal structure of a given molecular compound, and provides a set of low-energy metastable structures at a highly affordable cost.

The reasons why evolutionary algorithms succeed in crystal structure prediction have been discussed before \cite{Oganov-ACC}. As mentioned in Sec. II, in addition to these, the \emph{constrained global optimization} fixes the molecular connectivity, and brings the need for new variation operators (rotational mutation and softmutation), developed and described here.

For efficient and reliable polymorph prediction, the population of structures should be sufficiently diverse. A major difficulty in the prediction of molecular crystals is the large number of plausible candidate structures that can have very close energies \cite{Neumann:2005}. Given the complexity of their energy landscape, high diversity of the population of the structures is mandatory for successful prediction of molecular crystal structures. The initial population is particularly important, and it is usually a good idea to add a number of random symmetrized structures in each generation, to keep sampling of the landscape diverse.

The presented algorithm provides not only the theoretical ground state, but also a number of low-energy metastable structures. With inclusion of zero-point energy and entropic contributions, such structures may become stable. Even if this does not happen, low-energy metastable structures have a relatively high chance to be synthesized at special conditions.

While DFT+D is today's state of the art and its accuracy is often sufficient, for some systems (glycine), DFT+D is too crude, and more reliable approaches for computing the energy are needed. Under high pressure many of the difficulties disappear, because the vdW interactions (poorly accounted for by today's \emph{ab initio} methods) become relatively less important.

Clearly, the quality of the global minimum found by USPEX depends on the accuracy of the theory used for energy calculations and structure relaxation. Current levels of theory can be roughly divided into empirical, semiempirical, and \emph{ab initio} approaches. Accurate empirical force fields are appropriate for CSP, but reliable parameterizations are hard to generate for most molecules. In contrast to empirical force fields, \emph{ab initio} calculations provide a more accurate and rigorous description without parameterization, but the calculations are much more time-consuming. In our prediction, we adopt the DFT+D level of theory, which combines "the best of both worlds", i.e. an accurate representation of intermolecular repulsions, hydrogen bonding, electrostatic interactions, and vdW dispersions. DFT+D proved to be reliable for most systems, but its results are not fully satisfactory for glycine. This shows that further improvements in theoretical calculations of intermolecular interactions energies are needed. In parallel with the improvement of methods for energy ranking, there is a need for efficient and reliable algorithms for global optimization of the theoretical energy landscape, and present work is an important development in this direction. In the present paper, we describe the most important ingredients of this method, and demonstrate how it enables affordable structure prediction for many complex organic and inorganic systems at \emph{ab initio} level.

In summary, we have presented a new efficient and reliable approach for global energy optimization for molecular crystal structure prediction. It is based on the evolutionary algorithm USPEX extended to molecular crystals by additional variation operators and constraints of partially or completely fixed molecules. The high efficiency of this method enables fully quantum-mechanical structure predictions to be performed at an affordable computational cost. Using this method, we succeeded in finding the stable structures for systems with various rigid molecular shapes (tetrahedral, linear, bent, planar and complex molecules), and different bonding situations (vdW bonding, ionic, covalent, metallic, weak and strong hydrogen bonding, $\pi$-$\pi$ stacking, etc). We showed that even large systems can be efficiently dealt with by this approach, even at the \emph{ab initio} level of theory. This new approach also has wide applicability to inorganic crystals containing clusters and complex ions.

\section{Acknowledgments}
Calculations were performed on the supercomputer of Center for Functional Nanomaterials, Brookhaven National Laboratory, which is supported by the U.S. Department of Energy, Office of Basic Energy Sciences, under contract No. DE-AC02-98CH10086, and on Skif-MSU supercomputer (Moscow State University, Russia) and at the Joint Supercomputer Center (Russian Academy of Sciences, Moscow, Russia). This work is funded by DARPA (grant N66001-10-1-4037), National Science Foundation (grant EAR-1114313). We thank Prof. A. Garcia, Dr. S. E. Boulfelfel and Dr. J. Perez for insightful discussions.

\bibliography{1.bib}

\end{document}